\documentclass[aps,prb,preprint]{revtex4-1} 

\usepackage{amsfonts} % needed for bold Greek, Fraktur, and blackboard bold
\usepackage{graphicx} % needed for figures

\def\bea{\begin{eqnarray}}
\def\eea{\end{eqnarray}}
\def\be{\begin{equation}}
\def\ee{\end{equation}}

\begin{document}

\title{Klein--Gordon equation in curved space-time}

\author{Rebekah D. Lehn}
\author{Sophia S. Chabysheva}
\email{schabysh@d.umn.edu}
\author{John R. Hiller}
\affiliation{Department of Physics and Astronomy, University of Minnesota-Duluth, Duluth, MN 55812 USA}

\date{\today}

\begin{abstract}
We solve the relativistic Klein--Gordon equation for a light particle gravitationally
bound to a heavy central mass, with the gravitational interaction prescribed by the metric
of a spherically symmetric space-time.  Metrics are considered for an impenetrable 
sphere, a soft sphere of uniform density, and a soft sphere with a linear transition 
from constant to zero density; in each case the radius of the central
mass is chosen to be sufficient to avoid any event horizon.  The solutions
are obtained numerically and compared with nonrelativistic Coulomb-type solutions,
both directly and in perturbation theory, to study the general-relativistic 
corrections to the quantum solutions for a $1/r$ potential.  The density profile
with a linear transition is chosen to avoid singularities in the wave equation
that can be caused by a discontinuous derivative of the density.
\end{abstract}

\maketitle

\section{Introduction} \label{sec:introduction}

Although gravity is too weak for there to be, in practice, a gravitational analog
of the hydrogen atom,\cite{PhaseShift} the quantum mechanics of a particle bound 
gravitationally to a central mass can still be considered theoretically.  Of course, 
the nonrelativistic form of this problem is trivially solved, given that the
hydrogen-atom Coulomb solutions are so well known.  What is of some interest,
however, is how such a problem can be solved in the context of the general
theory of relativity, where the gravitational interaction is defined by
a space-time metric.  The formulation of the problem must then start from
a covariant relativistic wave equation, such as the Klein--Gordon (KG) equation
or the Dirac equation.  For simplicity, we consider the former.

The solution of the KG equation in a curved space-time requires
numerical techniques.  Earlier work that sought exact solutions\cite{exactKG}
was inconclusive at best, involving either approximate expansions or
simplifying assumptions to obtain asymptotic solutions.  On the
other hand, numerical methods, such as matrix methods for finite-difference
approximations and Runge-Kutta integration combined with boundary-condition
matching, are easy to implement to almost any desired accuracy.

Our models are based on static, spherically symmetric metrics of the form
\be
ds^2=g_{00}(r)dt^2-g_{rr}(r)dr^2-r^2d\theta^2-r^2\sin^2\theta d\phi^2.
\ee
The classic example is the Schwarzschild metric, for which,\cite{units} 
outside the central mass $M$,
\be \label{eq:g00grr}
g_{00}=1-\frac{2GM}{r},\;\;g_{rr}=\left(1-\frac{2GM}{r}\right)^{-1}.
\ee
For the impenetrable sphere, the
KG wave function is set to zero at the outer radius $r_0$ of the mass,
with $r_0$ always chosen larger than the Schwarzschild radius $r_S\equiv 2GM$.
For all spherically symmetric models, the Schwarzschild metric is
the exterior solution.
For the interior of the soft sphere, we use the solution\cite{softsphere}
for a uniform mass density of radius $r_0$
\be \label{eq:softmetric}
g_{00}=\frac14\left[3\sqrt{f(r_0)}-\sqrt{f(r)}\right]^2,\;\; g_{rr}=1/f(r),
\ee
with
\be
f(r)=1-\frac{r_S}{r_0^3}r^2.
\ee
For consistency of the model, the radius $r_0$ must be greater than $\frac98 r_S$;
this is known as the Buchdahl limit.\cite{limit}
As a tensor, the metric is then specified by the diagonal matrices
\be
g^{\mu\nu}=\left(\begin{array}{cccc} g_{00} & 0 & 0 & 0 \\
                                      0 & -g_{rr} & 0 & 0 \\
                                      0 & 0 & -r^2 & 0 \\
                                      0 & 0 & 0 & -r^2\sin^2\theta \end{array}\right),\;\;\;\;
g_{\mu\nu}=\left(\begin{array}{cccc} 1/g_{00} & 0 & 0 & 0 \\
                                      0 & -1/g_{rr} & 0 & 0 \\
                                      0 & 0 & -1/r^2 & 0 \\
                                      0 & 0 & 0 & -1/r^2\sin^2\theta \end{array}\right).
\ee

Given such a metric, the covariant KG equation $(\nabla^\mu g_{\mu\nu} \nabla^\nu +m^2)\Psi=0$
for a mass $m$ takes the form
\be
\left[\frac{1}{g_{00}}\frac{\partial^2}{\partial t^2}
   -\frac{1}{r^2}\frac{\partial}{\partial r}\left(\frac{r^2}{g_{rr}}\frac{\partial}{\partial r}\right)
   +\frac{L^2}{r^2}+m^2\right]\Psi=0,
\ee
with $L^2$ the operator for total angular momentum.
Standard separation of variables, as $\Psi=\tau(t)R_l(r)Y_{lm}(\theta,\phi)$ with $Y_{lm}$ the
usual spherical harmonics, yields
\be \label{eq:radialKG}
-\frac{1}{\tau}\frac{d^2\tau}{dt^2}
  =g_{00}\left[-\frac{1}{r^2R_l}\frac{d}{dr}\left(\frac{r^2}{g_{rr}}\frac{dR_l}{dr}\right)
  +\frac{l(l+1)}{r^2}+m^2\right]\equiv E^2,
\ee
where $E^2$ is the separation constant, chosen such that $E$ is the total energy
and $\Delta E\equiv E-m$, the binding energy.  The solution for $\tau$ is, of course,
trivial: $\tau=e^{\pm iEt}$, and we do not consider it further.  Our interest is in
the stationary states, with radial wave functions $R_l$, and their energy levels.\cite{positive}

The wave functions are normalized as integrals over the volume of the curved
space.  The volume of a sphere of radius $r_0$ is given by
\be
V=\int^{r<r_0} \sqrt{|g|}d^3r=4\pi\int_0^{r_0} \sqrt{g_{rr}}r^2dr,
\ee
where $|g|=g_{rr}r^4\sin^2\theta$ is the determinant of the spatial part of the metric.
Thus the normalization condition for the radial wave function is
\be
1=\int |R_l|^2 \sqrt{g_{rr}} r^2 dr.
\ee

In the following section, we discuss the solution of the radial equation for the two models
and separate
the relativistic corrections to be compared with perturbation theory.  In Sec.~\ref{sec:results}
we present results for the energies and wave functions, including comparisons with
perturbation theory and comparisons between models.  The results show that the sharp 
boundary between the interior and exterior of the soft sphere induces a kink in the
wave function; we therefore consider a modification of the model to taper the edge
in Sec.~\ref{sec:taper}.  A summary of the work and
possible extensions are discussed in Sec.~\ref{sec:summary}.  Details of one
derivation are left to an appendix.

\section{Analysis} \label{sec:analysis}

As a first step in analyzing the radial KG equation (\ref{eq:radialKG}),
we introduce a modified radial wave function $u_l$ such that no first-order
derivatives appear.  As shown in the Appendix, this is accomplished with
the definition $R_l=\sqrt{g_{rr}}u_l/r$.  The radial equation then reduces to
\be
-\frac{d^2u_l}{dr^2}-\left[\frac{g'_{rr}}{rg_{rr}}+\frac12\frac{g''_{rr}}{g_{rr}}
                           -\frac34\left(\frac{g'_{rr}}{g_{rr}}\right)^2\right]u_l
                           +\frac{l(l+1)}{r^2}g_{rr}u_l+m^2g_{rr}u_l=E^2\frac{g_{rr}}{g_{00}}u_l.
\ee
In the exterior region, where the metric is always the Schwarzschild metric, the
contents of the square brackets simplifies considerably.  The required derivatives
are
\be
g'_{rr}=- \frac{2GM}{r^2}\left(1-\frac{2GM}{r}\right)^{-2}=-r_S/(r-r_S)^2
\ee
and
\be
g''_{rr}=2r_S/(r-r_S)^3.
\ee
Substitution of these yields, for $r>r_0$,
\be
-\frac{d^2u_l}{dr^2}-\frac14\frac{r_S^2}{r^2(r-r_S)^2}u_l
                           +\frac{l(l+1)}{r^2(1-r_s/r)}u_l+\frac{m^2}{1-r_s/r}u_l=\frac{E^2}{(1-r_s/r)^2}u_l.
\ee

As discussed more fully below, the leading Coulombic interaction arises not from the second
term but from the difference of the $m^2$ and $E^2$ terms.  Here $E$ is the relativistic
energy, which includes the rest energy $m$; a nonrelativistic Schr\"odinger equation
would consider the difference, $\Delta E\equiv E-m$.  Substitution of the expansion
$E^2=m^2+2m\Delta E+\Delta E^2$, combination of the two terms proportional to $m^2$, 
and division by $2m$ leads to
\be \label{eq:reduced}
-\frac{1}{2m}\frac{d^2u_l}{dr^2}-\frac14\frac{r_S^2}{2mr^2(r-r_S)^2}u_l
                           +\frac{l(l+1)}{2mr^2(1-r_S/r)}u_l
                -\frac{mr_S}{2r}\frac{1}{(1-r_S/r)^2}u_l
                =\frac{\Delta E+\Delta E^2/2m}{(1-r_S/r)^2}u_l.
\ee
The combination $mr_S/2$ is just $GMm$.  Thus to lowest order in $r_S$, the
modified radial equation reduces to
\be
-\frac{1}{2m}\frac{d^2u_l}{dr^2}+\frac{l(l+1)}{2mr^2}u_l-\frac{GMm}{r}u_l=\Delta E u_l,
\ee
which is just the standard Schr\"odinger equation with a Newtonian gravitational
potential.  We discuss leading corrections to this below.

Given that $GMm$ plays the role of $e^2$ in the Coulomb term, the natural length scale 
for this system is the Bohr radius $a=1/GMm^2$.  Consequently, we rescale
the radial coordinate to a dimensionless variable $\zeta=r/a$ and correspondingly
rescale the Schwarzschild radius as $\zeta_S=r_S/a=2(GMm)^2$ and the sphere radius
as $\zeta_0=r_0/a$.  The natural energy scale
is $GMm/2a$, which leads to a dimensionless binding energy $\epsilon\equiv 2a\Delta E/GMm$.
In terms of these dimensionless quantities, the modified radial equation becomes
\be \label{eq:dimensionless}
-\frac{d^2u_l}{d\zeta^2}-\left[\frac{g'_{rr}}{\zeta g_{rr}}+\frac12\frac{g''_{rr}}{g_{rr}}
                           -\frac34\left(\frac{g'_{rr}}{g_{rr}}\right)^2\right]u_l
                           +\frac{l(l+1)}{\zeta^2}g_{rr}u_l+m^2g_{rr}(1-1/g_{00})u_l=(\epsilon+\frac18\zeta_S\epsilon^2)\frac{g_{rr}}{g_{00}}u_l,
\ee
with primes now defined to mean differentiation with respect to $\zeta$.  It is
this equation that we solve numerically.

For ordinary gravity, the effects are extremely small.  For an electron bound to
a proton, the gravitational fine structure constant $GMm$ is just $3.2\times10^{-42}$,
and the dimensionless Schwarzschild radius is $2\times10^{-83}$.
These small numbers mean that any numerical solution will not be able to
resolve any relativistic effects.  Only the Coulombic binding energies $\epsilon=-1/n^2$
would be calculable.  To have meaningful calculations we must assume a much stronger
gravitational coupling and consider values of $\zeta_S$ no more than a few orders of magnitude
less than unity.

We used two methods to solve the modified radial equation, in order to have some
basis for checking the work.  One method was to integrate the equation both outward
and inward to a point near one Bohr radius, at which the log derivative of the wave 
function was required to match between the two integrations; the values of epsilon
for which a match was achieved were the eigenvalues.  The integrations were done
with an adaptive Runge--Kutta--Fehlberg algorithm\cite{DeVries} for a system of first order
equations equivalent to the given second order equation.  

The other method of solution was to apply a simple finite-difference approximation to 
the equation and thereby convert it to a matrix equation where the eigenvalues of the matrix
correspond to $\lambda\equiv\epsilon+\frac18\zeta_S\epsilon^2$ and the eigenvectors determine
the wave function at the discrete points used for the finite differences.
The actual $\epsilon$ values are obtained by solving the quadratic equation
implied by the definition of $\lambda$, which leaves
\be
\epsilon=\frac{-1\pm\sqrt{1+\zeta_S\lambda/2}}{\zeta_S/4}.
\ee
To be consistent with the limit that $\epsilon$ equal $\lambda$ when $\zeta_S$ goes
to zero, the upper sign is chosen, and to facilitate this limit computationally,
we rearrange the expression as
\be
\epsilon=\frac{2\lambda}{1+\sqrt{1+\zeta_S\lambda/2}}
\ee
to avoid the indeterminant $0/0$.  The accuracy of the eigenvalues is
improved by Richardson extrapolation\cite{DeVries} from a set of different grid spacings.

As another check on the calculations, we consider first-order perturbation theory
for the leading relativistic contributions for the case of the impenetrable sphere.
From (\ref{eq:reduced}) or (\ref{eq:dimensionless}), the dimensionless form of the radial equation can
be seen to be
\be 
-\frac{d^2u_l}{d\zeta^2}-\frac14\frac{\zeta_S^2}{\zeta^2(\zeta-\zeta_S)^2}u_l
                           +\frac{l(l+1)}{\zeta^2(1-\zeta_S/\zeta)}u_l
                -\frac{2}{\zeta}\frac{1}{(1-\zeta_S/\zeta)^2}u_l
                =\frac{\epsilon+\frac18\zeta_S\epsilon^2}{(1-\zeta_S/\zeta)^2}u_l.
\ee
Keeping $\zeta_S$ to first order, we obtain
\be
-\frac{d^2u_l}{d\zeta^2}+\frac{l(l+1)}{\zeta^2}(1+\zeta_S/\zeta)u_l
                -\frac{2}{\zeta}(1+2\zeta_S/\zeta)u_l
                =(\epsilon(1+2\zeta_S/\zeta)+\frac18\zeta_S\epsilon^2)u_l.
\ee
If we collect all of the ${\cal O}(\zeta_S)$ corrections into a 
perturbing potential, with $\epsilon$ replaced by its leading value $-1/n^2$,
\be
V_S(\zeta)=\zeta_S\left[\frac{l(l+1)}{\zeta^3}-\frac{4}{\zeta^2}
            +\frac{1}{n^2}\frac{2}{\zeta}-\frac18\frac{1}{n^4}\right],
\ee
the radial equation with this first order correction reads
\be
-\frac{d^2u_l}{d\zeta^2}+\frac{l(l+1)}{\zeta^2}u_l
                -\frac{2}{\zeta}u_l+V_Su_l=\epsilon u_l.
\ee
If we keep the central radius $r_0$ small and ignore the small deviation 
from a pure Coulomb interaction inside $r_0$, where the zero-order 
potential is infinite, the zero-order part of this equation yields $-1/n^2$ 
as the zero-order eigenvalue, and the shift due to $V_S$ is 
$\langle u_l|V_S|u_l\rangle$, with $u_l$ approximated by the standard
hydrogenic modified radial wave functions.  For these wave functions,
the expectation values of powers of $\zeta$ are known.  We need
\be
\langle\frac{1}{\zeta}\rangle=\frac{1}{n^2},\;\;
\langle\frac{1}{\zeta^2}\rangle=\frac{1}{(l+1/2)n^3},\;\;
\langle\frac{1}{\zeta^3}\rangle=\frac{1}{(l(l+1)(l+1/2)n^3}.
\ee
On substitution, they provide
\be \label{eq:perttheory}
\langle V_S\rangle=\frac{\zeta_S}{(l+1/2)n^4}\left[\frac{15}{8}(l+1/2)-3n\right].
\ee
This can be compared to shifts in the numerical eigenvalues of the full radial equation
as $\zeta_S$ is varied.
The approximations made are very good for nonzero angular momentum states,
for which the wave function does not significantly explore the small-$r$
region, due to the repulsive $1/r^2$ term.

\section{Results} \label{sec:results}

The comparison with perturbation theory is shown in Table~\ref{tab:perttheory},
where the coefficient of $\zeta_S$ in $\langle V_S\rangle$ is extracted
as the slope of a least-squares fit to a line.
Here we see that the predictions for $l=1$ and $l=2$ are quite close.
For $l=0$ the
agreement is not very good, but this is due to the corrections that
should be made for small radii, where the Coulombic solution, used
to compute the energy shift, is a poor approximation.

\begin{table}[ht]
\centering
\caption{Comparison with perturbation theory for the lowest energy 
levels, $\epsilon_{l+1}$, for the impenetrable sphere. The slopes
relative to $\zeta_S$, and the associated errors, were computed by a 
least-squares fit.  The theory estimate is given by 
Eq.~(\ref{eq:perttheory}) of the text.}
\begin{ruledtabular}
\begin{tabular}{cc|ccc}
$\zeta_S$ & $\zeta_0$ & $l=0$ & $l=1$ & $l=2$ \\
\hline
0 & 0 & -1 & -0.25 & -0.11111 \\
0.005 & 0.0051 & -1.0122 & -0.25067 & -0.11122 \\
0.010 & 0.011  & -1.0223 & -0.25134 & -0.11132 \\
0.015 & 0.016  & -1.0383 & -0.25203 & -0.11143 \\
0.020 & 0.021  & -1.0558 & -0.25272 & -0.11154 \\
0.030 & 0.031  & -1.0951 & -0.25414 & -0.11176 \\
\hline
\multicolumn{2}{c|}{slope} & -3.37$\pm$0.21 & -0.1389$\pm$0.0009 & -0.0217$\pm$0.0002 \\
\multicolumn{2}{c|}{theory estimate} & -4.125 & -0.1328 & -0.0213
\end{tabular}
\end{ruledtabular}
\label{tab:perttheory}
\end{table}

Tables~\ref{tab:hardsphere} and \ref{tab:softsphere} list some binding energies
computed for each model for similar values of the parameters, $\zeta_0$ and $\zeta_S$.
The radius of the sphere $\zeta_0$ is held fixed at 0.5, which corresponds to 1/2
of a Bohr radius.  The Schwarzschild radius $\zeta_S$, which parameterizes the
strength of the relativistic effects, is varied.  The effect on the ground
state is quite striking, particularly for the soft sphere.  In contrast,
the lowest $l=2$ state remains almost unaffected, with all values very close
to the Coulombic $-1/9$.  Between the two models, the soft sphere has the
more deeply bound states; the repulsive hard sphere acts to push the particle
outward and out of the deeper regions of the effective potential.

\begin{table}[ht]
\centering
\caption{Dimensionless binding energies for an impenetrable sphere of radius $\zeta_0=0.5$.}
\begin{ruledtabular}
\begin{tabular}{c|ccc|cc|c}
         & \multicolumn{3}{c|}{$l=0$} & \multicolumn{2}{c|}{$l=1$} & $l=2$ \\
$\zeta_S$ & $\epsilon_1$ & $\epsilon_2$ &$\epsilon_3$ &$\epsilon_2$ &$\epsilon_3$ &$\epsilon_3$  \\
\hline
0    & -0.4891 & -0.1683 & -0.0845 & -0.2431 & -0.1087 & -0.1111 \\
0.01 & -0.4933 & -0.1695 & -0.0850 & -0.2443 & -0.1092 & -0.1113 \\
0.10 & -0.5369 & -0.1813 & -0.0896 & -0.2565 & -0.1139 & -0.1133 \\
0.20 & -0.6042 & -0.1992 & -0.0965 & -0.2737 & -0.1204 & -0.1158 \\
0.30 & -0.7112 & -0.2269 & -0.1067 & -0.2981 & -0.1296 & -0.1184 \\
0.40 & -0.9339 & -0.2836 & -0.1265 & -0.3427 & -0.1463 & -0.1214 
\end{tabular}
\end{ruledtabular}
\label{tab:hardsphere}
\end{table}

\begin{table}[ht]
\centering
\caption{Same as Table~\ref{tab:hardsphere} but for the soft-sphere model.}
\begin{ruledtabular}
\begin{tabular}{c|ccc|cc|c}
         & \multicolumn{3}{c|}{$l=0$} & \multicolumn{2}{c|}{$l=1$} & $l=2$ \\
$\zeta_S$ & $\epsilon_1$ & $\epsilon_2$ &$\epsilon_3$ &$\epsilon_2$ &$\epsilon_3$ &$\epsilon_3$  \\
\hline
0    & -0.899 & -0.237 & -0.107 & -0.250 & -0.1111 & -0.1111\\
0.01 & -0.936 & -0.243 & -0.109 & -0.251 & -0.1116 & -0.1113 \\
0.10 & -1.517 & -0.319 & -0.132 & -0.266 & -0.1173 & -0.1133 \\
0.20 & -2.970 & -0.453 & -0.167 & -0.293 & -0.1273 & -0.1158 \\
0.30 & -5.256 & -0.651 & -0.218 & -0.542 & -0.2148 & -0.1184 \\
0.38 & -7.701 & -1.703 & -0.543 & -3.157 & -0.3246 & -0.1208 \\
0.40 & -8.646 & -3.001 & -0.769 & -4.158 & -0.4202 & -0.2619
\end{tabular}
\end{ruledtabular}
\label{tab:softsphere}
\end{table}

\begin{figure}[ht]
\centering
\includegraphics[width=15cm]{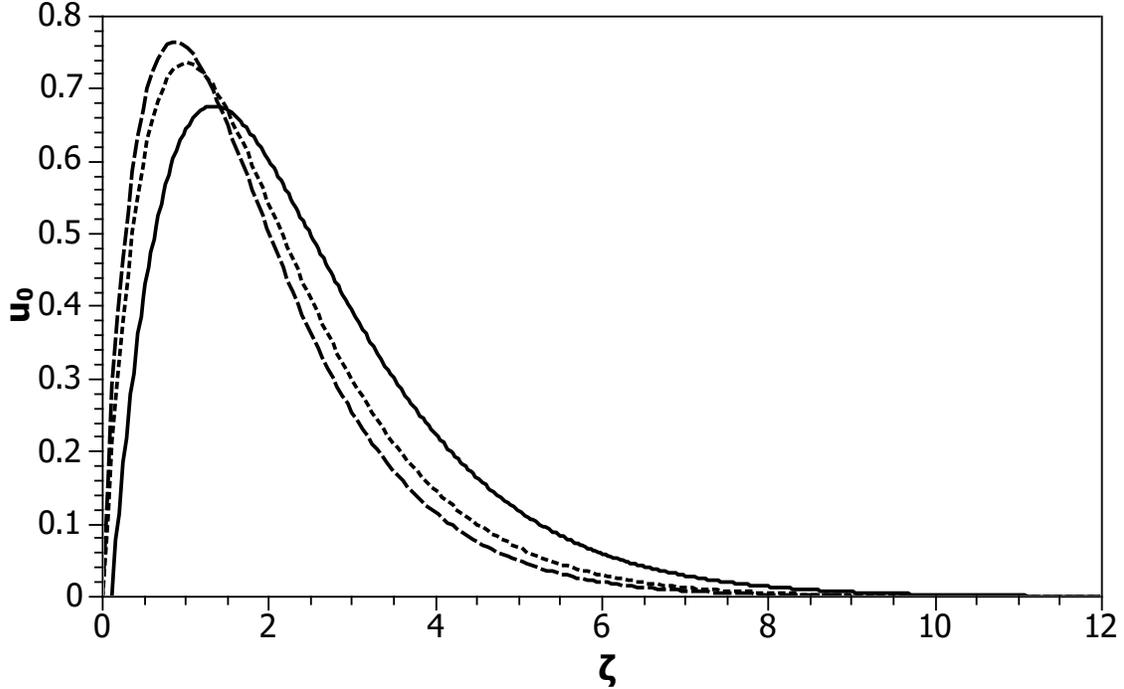}
\caption{Comparison of wave functions $u_0$ for the impenetrable sphere (solid), soft sphere (dashed)
and nonrelativistic Coulomb limit (dotted) with $\zeta_0=0.1$ as the radius of each sphere
and $\zeta_S=0.01$ as the Schwarzschild radius, all for zero angular momentum.  The radial 
coordinate $\zeta$ is scaled by the Bohr radius.}
\label{fig:hsc}
\end{figure}

\begin{figure}[ht]
\centering
\includegraphics[width=15cm]{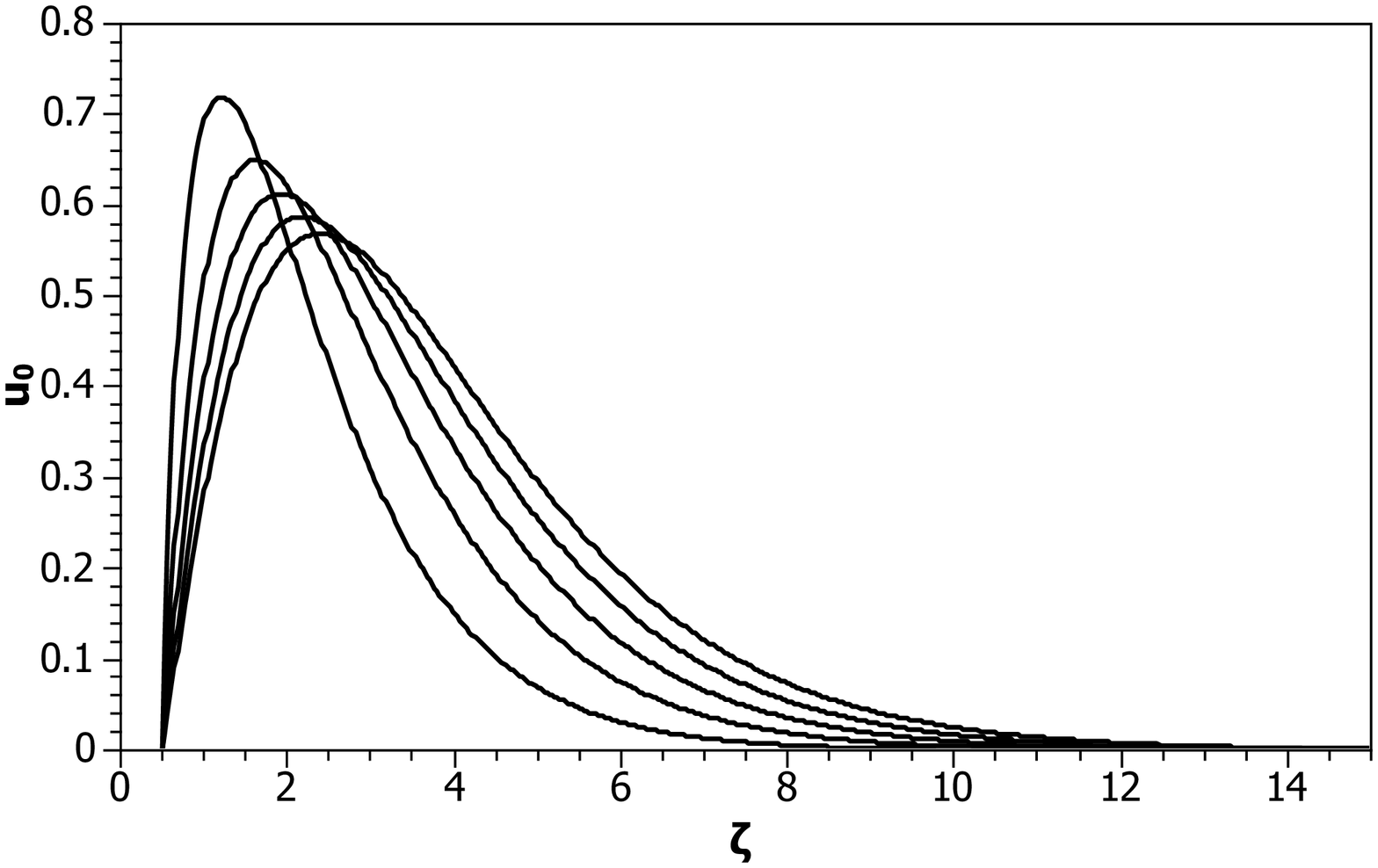}
\caption{Wave functions $u_0$ for an impenetrable sphere of radius $\zeta_0=0.5$ with
varying Schwarzschild radii $\zeta_S=0$, 0.1, 0.2, 0.3, and 0.4 and zero angular
momentum.  As $\zeta_S$ increases, the peak moves to the left.}
\label{fig:hard}
\end{figure}

\begin{figure}[ht]
\centering
\includegraphics[width=15cm]{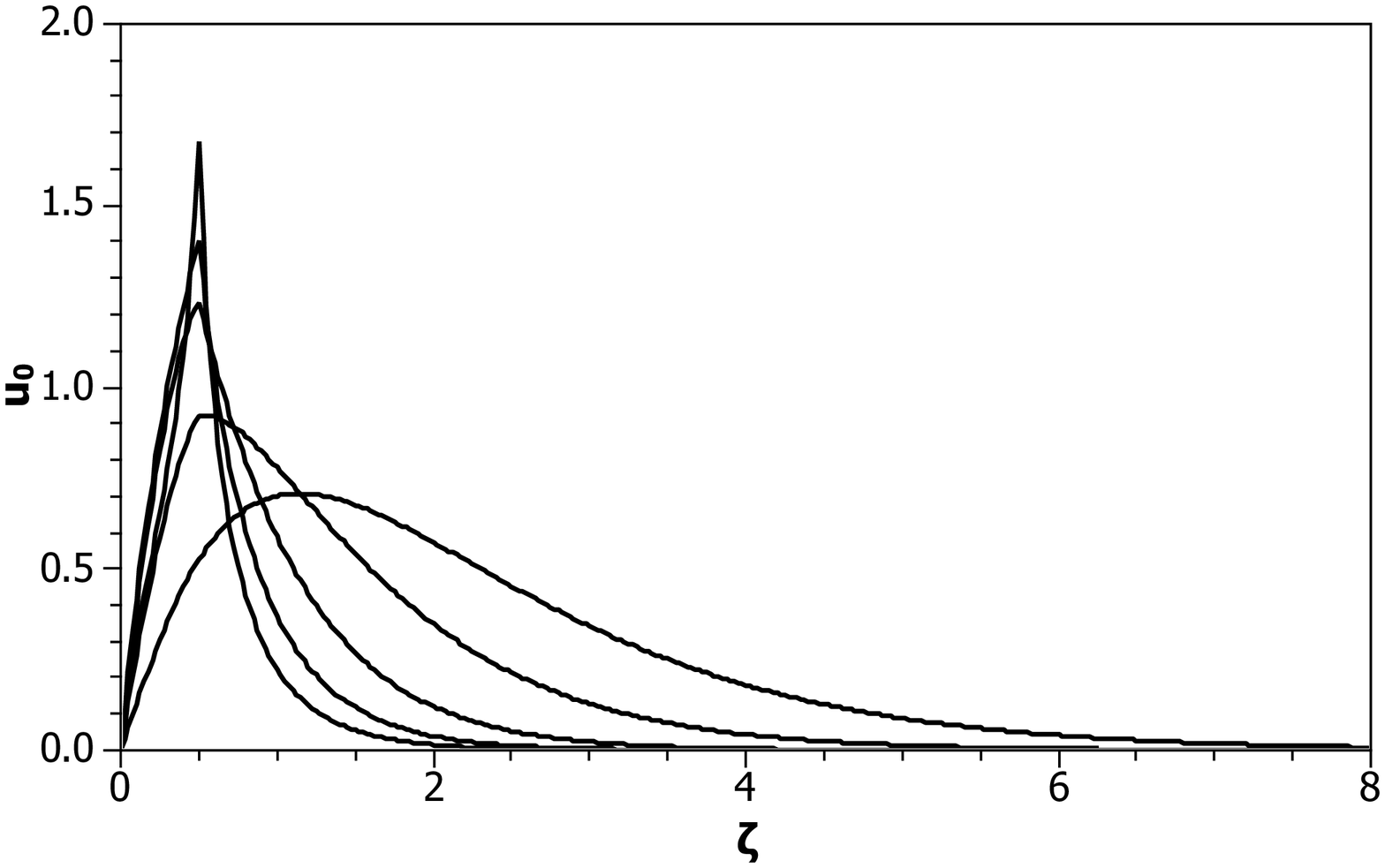}
\caption{Same as Fig.~\ref{fig:hard} but for the soft sphere.  As $\zeta_S$ increases,
the peak becomes sharper. The discontinuity
in the derivative at the sphere radius is due to discontinuities in the derivatives
of the metric.}
\label{fig:soft}
\end{figure}

\begin{figure}[ht]
\centering
\includegraphics[width=15cm]{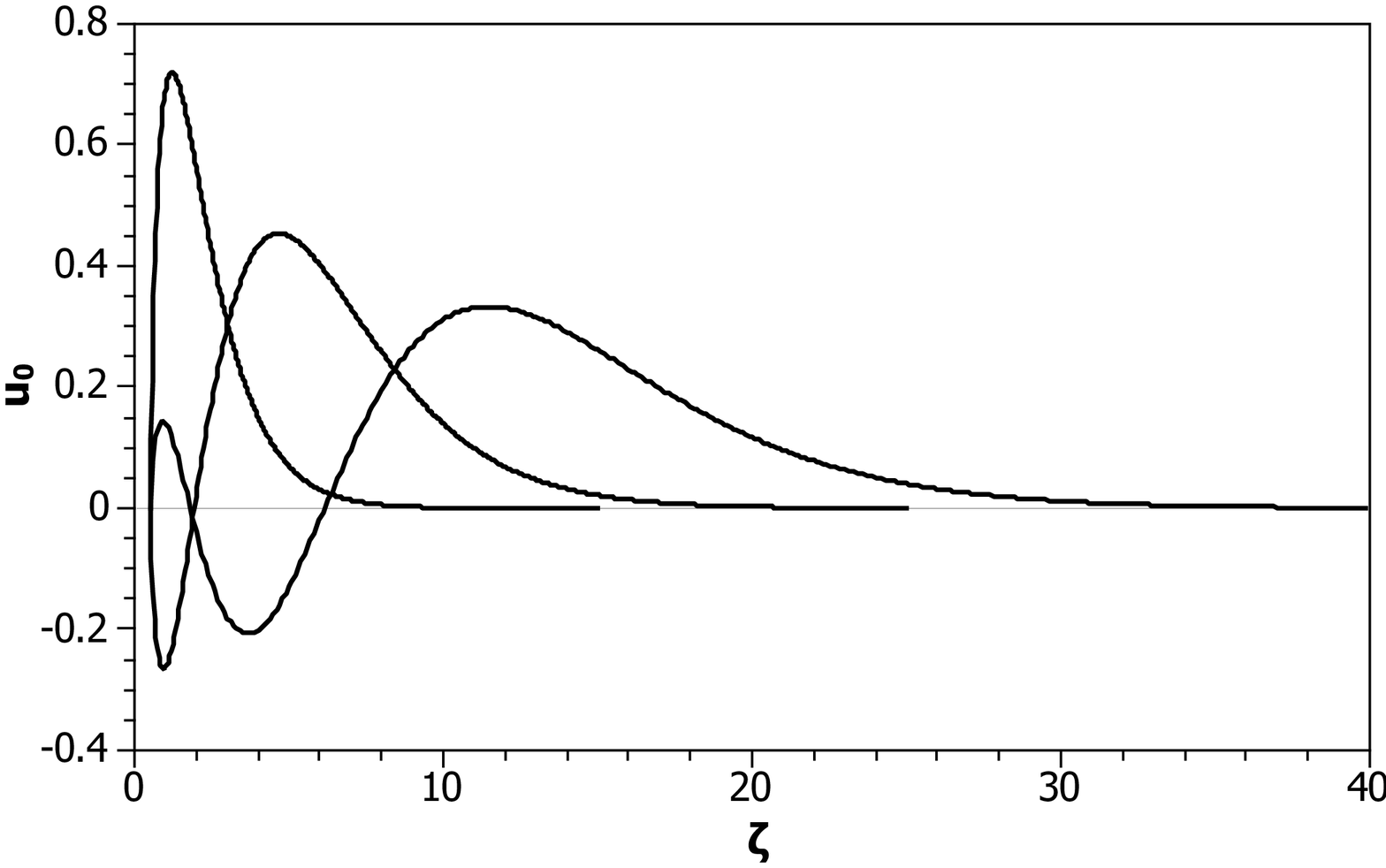}
\caption{Wave functions $u_0$ for three lowest levels of an impenetrable sphere of 
radius $\zeta_0=0.5$ with Schwarzschild radius $\zeta_S=0.4$ and zero angular momentum.}
\label{fig:hardn1to3}
\end{figure}

\begin{figure}[ht]
\centering
\includegraphics[width=15cm]{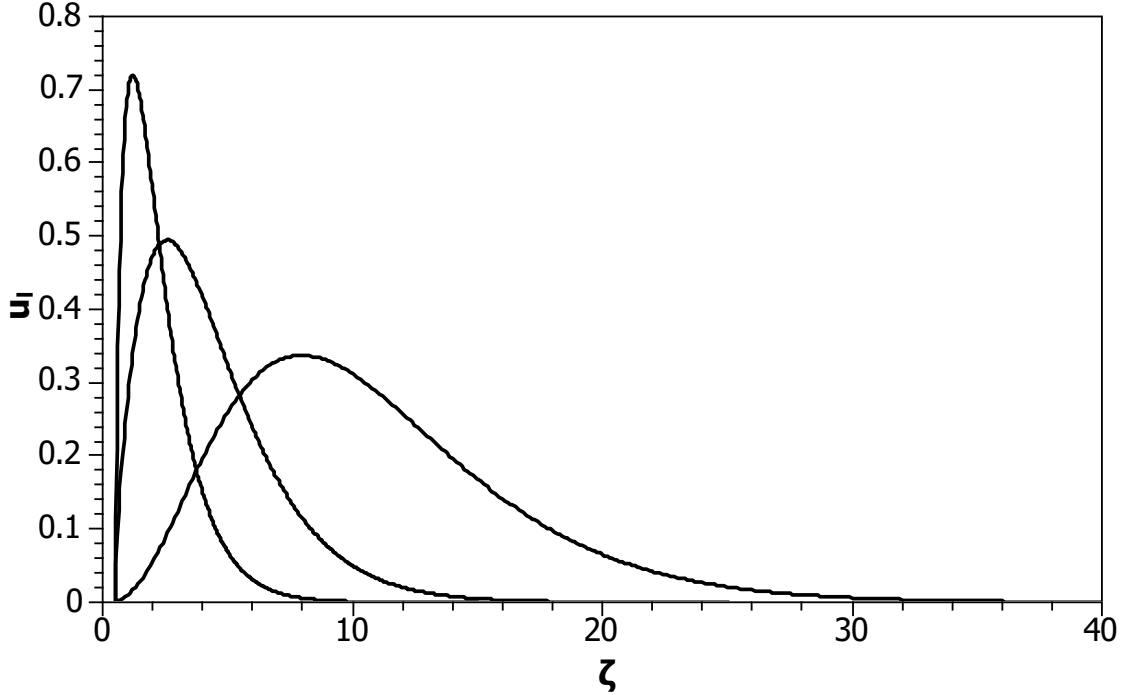}
\caption{Wave functions $u_l$ for the lowest level of an impenetrable sphere of 
radius $\zeta_0=0.5$ with Schwarzschild radius $\zeta_S=0.4$ and angular momentum $l=0,1,2$.}
\label{fig:hardl0to2}
\end{figure}

\begin{figure}[ht]
\centering
\includegraphics[width=15cm]{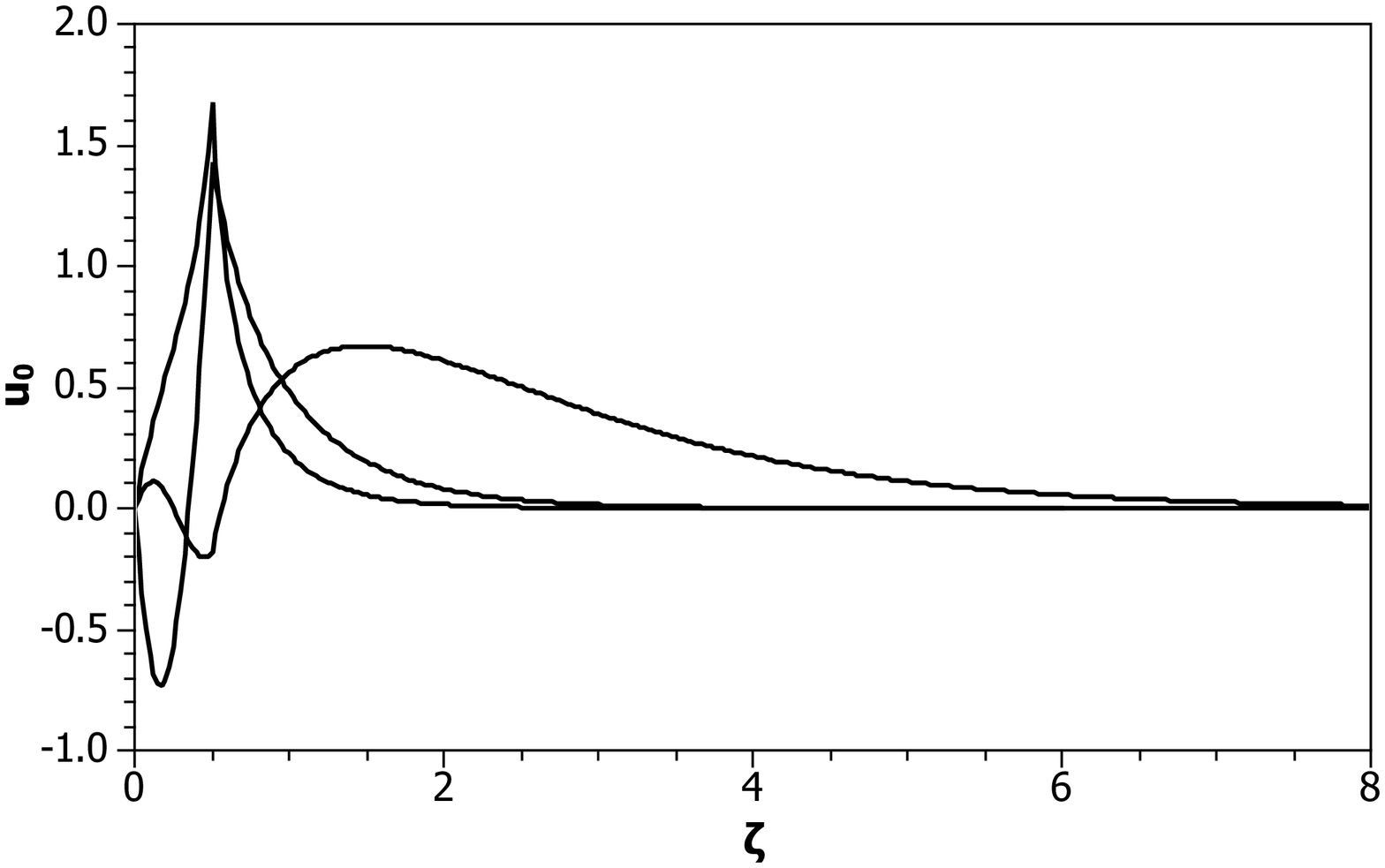}
\caption{Same as Fig.~\ref{fig:hardn1to3} but for the soft sphere.}
\label{fig:softn1to3}
\end{figure}

\begin{figure}[ht]
\centering
\includegraphics[width=15cm]{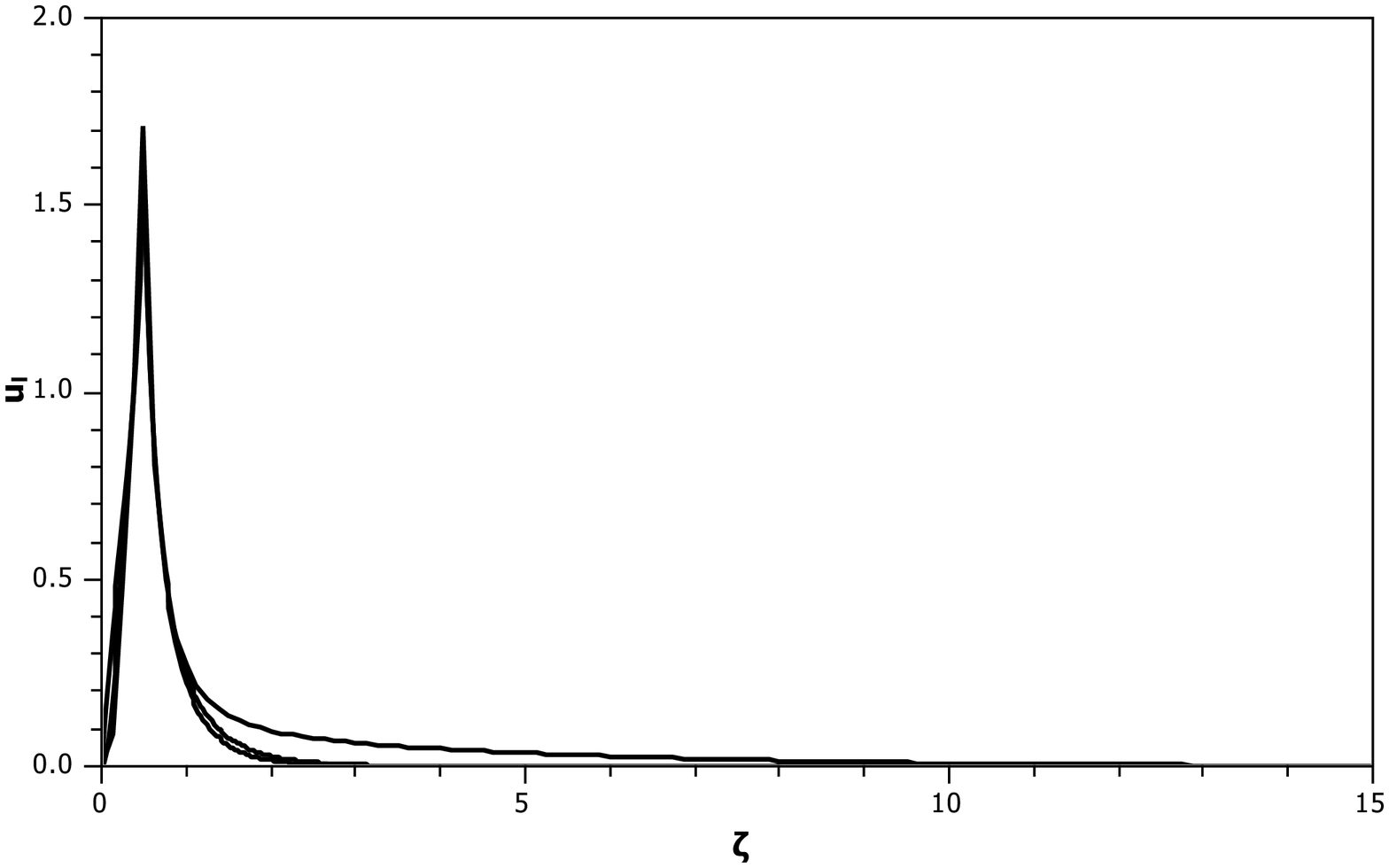}
\caption{Same as Fig.~\ref{fig:hardl0to2} but for the soft sphere.}
\label{fig:softl0to2}
\end{figure}

Some of the corresponding wave functions are plotted in Figs.~\ref{fig:hsc}-\ref{fig:softl0to2}.
Figure~\ref{fig:hsc} compares the wave functions for the two models with the standard
Coulomb solution, for a sphere with a small radius and a very small Schwarzschild radius.
The most significant difference is between the impenetrable sphere and the other two cases,
where the wave function for the impenetrable sphere is pushed outward relative to the other two.

Figure~\ref{fig:hard} shows the evolution of the impenetrable-sphere wave function as
the relativistic effects are increased.  As the Schwarzschild radius increases, the
wave function moves in, closer to the sphere.  The same process takes place for
the soft-sphere model, as shown in Fig.~\ref{fig:soft}.  Here, however, we see that
the discontinuity in the derivative of the metric that occurs at the sphere boundary,
at $\zeta=\zeta_0$, does induce a kink in the wave function; for the highly
relativistic cases, with larger $\zeta_S$ values, the kink becomes a sharp peak at
the sphere radius.

Comparisons of radially and rotationally excited states can be found in 
Figs.~\ref{fig:hardn1to3}-\ref{fig:softl0to2}.  The value of $\zeta_S$ is
such that the model is highly relativistic.  The radially excited states
show the standard nodal structure, and the different angular momentum
states show the effect of exclusion by the $l(l+1)/r^2$ contribution to the
effective potential.  The kink in the soft-sphere wave functions at the
sphere boundary becomes less pronounced for radially excited states,
for which the greater probability has moved away from the sphere.

\section{Continuous density function} \label{sec:taper}

\begin{figure}[ht]
\centering
\includegraphics[width=15cm]{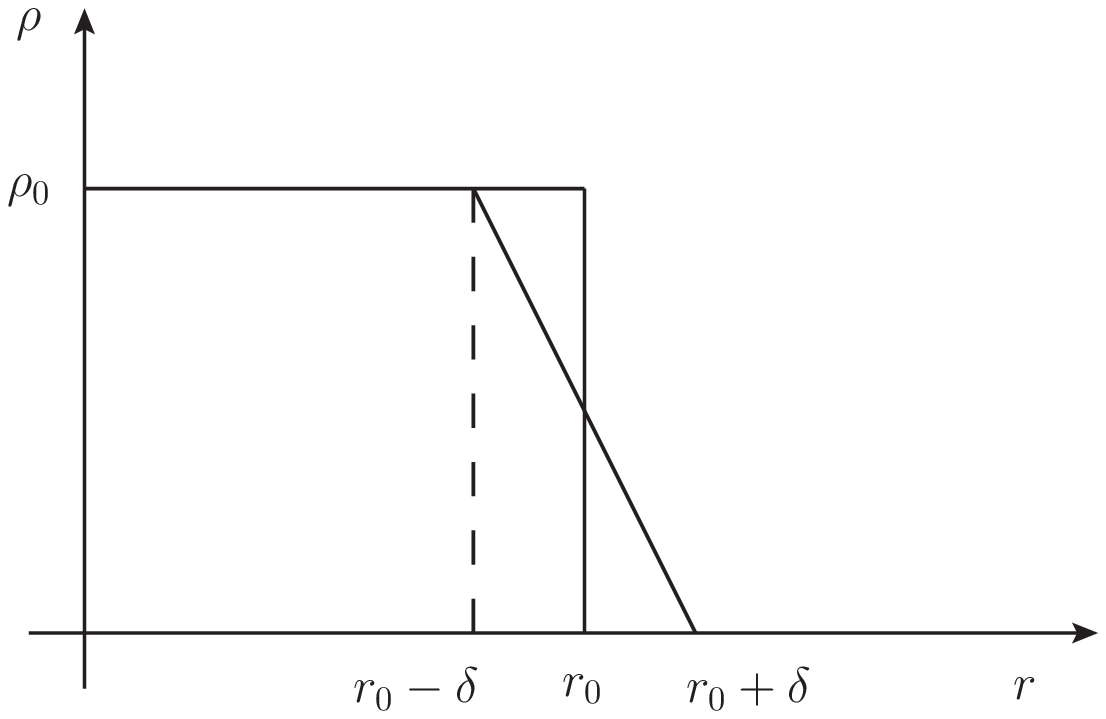}
\caption{Mass density model with a gradual transition to zero compared to
the discontinuous form with a sharp boundary at $r_0$.}
\label{fig:density}
\end{figure}

To avoid the sharp discontinuity in the density $\rho$, we
consider a more realistic behavior that models a gradual
transition with the simple form
\be
\rho(r)=\left\{\begin{array}{ll} \rho_0, & r\leq r_0-\delta \\ 
                                 (r-r_0-\delta)/(2\delta), & r_0-\delta<r\leq r_0+\delta \\
                                 0, & r_0+\delta<r \end{array}\right.
\ee
plotted in Fig.~\ref{fig:density}.
The parameter $\delta$ controls the width of the transition
and $\rho_0$, the normalization.  The mass inside a spherical
surface of area $4\pi r^2$ is
\be
\mu(r)=\int_0^r 4\pi r^2 \rho(r) dr,
\ee
with the total mass being $M=\mu(\infty)$; this fixes $\rho_0$
in proportion to $M$.  For the model, the integral is easily
done analytically.

Next, we construct the metric corresponding to this density profile.
The radial part of the metric is determined as\cite{exact}
\be
g_{rr}=[ 1-2G\mu(r)/r]^{-1}.
\ee
At large $r$, this reduces to the Schwarzschild expression, given in 
(\ref{eq:g00grr}).
The time component $g_{00}\equiv e^{2\Phi(r)}$ is determined
implicitly by the differential equation\cite{exact}
\be
\frac{d\Phi}{dr}=G\frac{\mu(r)+4\pi r^3 p(r)}{r(r-2 G\mu(r))}
\ee
and the boundary condition $\Phi(r)\sim\ln\sqrt{1-2G\mu(r)/r}$ for
large $r$, again from the Schwarzschild metric (\ref{eq:g00grr}).
Here $p(r)$ is the pressure, which is determined by the 
Tolman--Oppenheimer--Volkov (TOV) equation\cite{exact}
\be
\frac{dp}{dr}=-G\frac{(\rho(r)+p(r))(\mu(r)+4\pi r^3 p(r))}{r(r-2G\mu(r))}
\ee
and the boundary condition $p(\infty)=0$.  This differential
equation must be solved first, so that $p(r)$ is available
for use in the equation for $\Phi$.

In terms of the dimensionless radial coordinate $\zeta=r/a$
and Schwarzschild radius $\zeta_S=2GM/a$, these equations
reduce to
\bea
\tilde\mu(\zeta)&=&3\int_0^\zeta \zeta^2\tilde\rho d\zeta, \\
\frac{d\Phi}{d\zeta}&=&\frac{\zeta_S}{2}\frac{\tilde\mu(\zeta)+3\zeta^3\tilde p(\zeta)}
                                       {\zeta[\zeta-\zeta_S\tilde\mu(\zeta)]}, \\
\frac{d\tilde p}{d\zeta}&=&-\frac{\zeta_S}{2}
   \frac{[\tilde\rho(\zeta)+\tilde p(\zeta)][\tilde\mu(\zeta)+3\zeta^3\tilde p(\zeta)]}
        {\zeta[\zeta-\zeta_S\tilde\mu(\zeta)]},
\eea
with $\tilde\mu\equiv \mu/M$, $\tilde\rho\equiv\frac{4\pi a^3}{3M}\rho$, and
$\tilde p \equiv \frac{4 \pi a^3}{3M}p$.  The boundary conditions become
\be
\tilde\mu(\infty)=1,\;\;
\Phi(\zeta)\sim\ln\sqrt{1-\zeta_S\tilde\mu/\zeta},\;\;
\tilde p(\infty)=0.
\ee
For computational purposes, we model these conditions as occurring at a finite
$\zeta_{\rm max}\gg\zeta_0$:
\be
\tilde\mu(\zeta_{\rm max})=1,\;\;
\Phi(\zeta_{\rm max})=\ln\sqrt{1-\zeta_S/\zeta_{\rm max}}, \;\;
\tilde p(\zeta_{\rm max})=0.
\ee
The differential equations are solved by the RK2 Modified Euler method,\cite{DeVries}
in order to have a discretization error consistent with the finite-difference
approximation used to solve the radial wave equation (\ref{eq:dimensionless}).
The equation for $\tilde\mu$ is integrated outward from zero, and the solution
renormalized to match the boundary condition.
The equations for $\Phi$ and $\tilde p$ are integrated inward from $\zeta_{\rm max}$.
In the limit that $\delta$ goes to zero, the analytic solution (\ref{eq:softmetric})
for the metric is obtained, as a check on the calculations.

As input to the radial wave equation, we need derivatives of the metric function 
$g_{rr}=[1-\zeta_S\tilde\mu/\zeta]^{-1}$.  With use of $d\tilde\mu/d\zeta=3\zeta^2\rho$,
we obtain
\be
g'_{rr}=\zeta_S\frac{3\zeta^3\tilde\rho-\tilde\mu}{(\zeta-\zeta_S\tilde\mu)^2}
\ee
and
\be
g''_{rr}=\frac{2\zeta_S^2(\tilde\mu/\zeta^2-3\zeta\tilde\rho)^2}{(1-\zeta_S\tilde\mu/\zeta)^3}
  +\frac{\zeta_S(2\tilde\mu/\zeta^3+3\zeta d\tilde\rho/d\zeta)}{(1-\zeta_S\tilde\mu/\zeta)^2}.
\ee
For our model, $d\tilde\rho/d\zeta$ is computed analytically.  

\begin{figure}[ht]
\centering
\includegraphics[width=15cm]{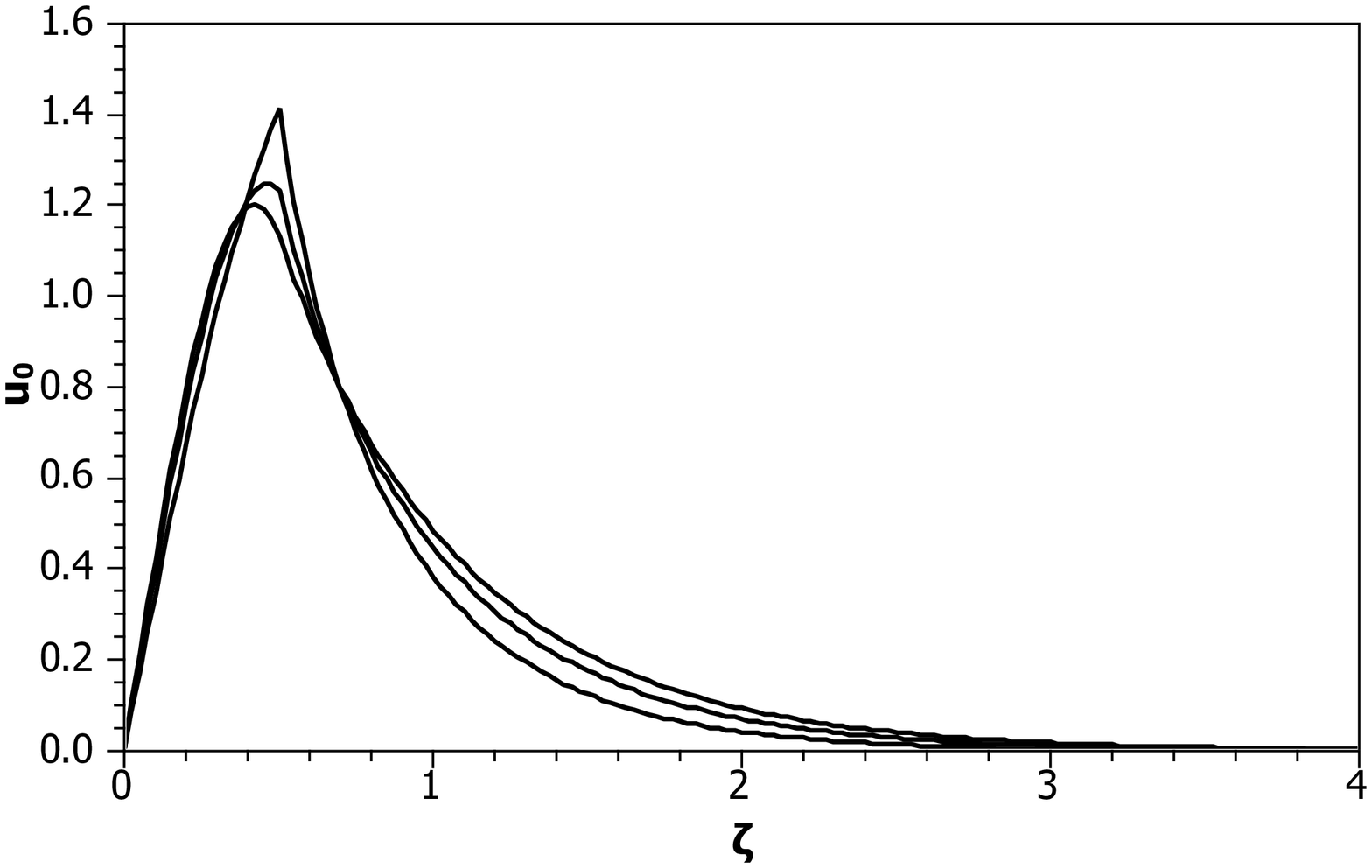}
\caption{Wave functions $u_0$ for the lowest level for a sphere with a gradual
transition to zero density, with nominal radius $\zeta_0=0.5$, Schwarzschild radius $\zeta_S=0.4$,
and zero angular momentum.  The transition parameter $\delta$ takes three values,
$0.05a$, $0.005a$ and 0, the last being the sharp transition with a kink in the wave 
function.  The wave function for the smaller nonzero value has the higher, rounded peak.}
\label{fig:vsphere}
\end{figure}

A sampling of the results is given in Table~\ref{tab:vsphere} and Fig.~\ref{fig:vsphere}.
The table shows that the gradual transition has a very immediate effect on the
ground state eigenenergy.  The entry for $\delta=0$ is just the previous result
for the sharp boundary.  For all nonzero values of $\delta$ the binding energy
is significantly less.  The wave functions are also greatly altered in the region
of the transition.  As can be seen in Fig.~\ref{fig:vsphere}, the kink is eliminated
and the peak nicely rounded when $\delta$ is nonzero.

\begin{table}[ht]
\centering
\caption{Dimensionless binding energies for a sphere of nominal radius $\zeta_0=0.5$
with a linear transition from constant density inside $\zeta_0-\delta/a$
to zero outside $\zeta_0+\delta/a$.  Results are given only for $S$ states and for
a dimensionless Schwarzschild radius of $\zeta_S=0.3$.}
\begin{ruledtabular}
\begin{tabular}{c|ccccc}
$\delta/a$ & 0 & 0.005 & 0.01 & 0.02 & 0.05 \\
\hline
$\epsilon_1$ & -5.256 & -3.196 & -3.192 & -3.182 & -3.099 
\end{tabular}
\end{ruledtabular}
\label{tab:vsphere}
\end{table}

\section{Summary} \label{sec:summary}

We have shown that solutions of the KG equation in curved space-time
can be computed with ordinary numerical methods and that the results
are consistent with the nonrelativistic limit.  The results for
binding energies are summarized in Tables~\ref{tab:hardsphere}, \ref{tab:softsphere},
and \ref{tab:vsphere}.  The scale of relativistic effects is set
by the dimensionless Schwarzschild radius $\zeta_S=2GM/a=(2GMm)^2$.
The binding becomes much stronger as $\zeta_S$ is increased, particularly
for the soft sphere, where the particle can penetrate the region of
nonzero mass density.  The addition of a linear transition reduces this
effect, primarily because the discontinuity in the metric derivative
created an artificially strong binding at the sharp boundary of the
soft sphere.  The associated wave functions are compared in a series
of figures, including Fig.~\ref{fig:vsphere} which shows the distinction
between soft spheres with and without the linear transition in density.

This work can be extended in at least two ways that would make nice
projects for advanced undergraduates and beginning graduate students.
One is to consider more sophisticated density profiles.  The
developments presented here can be immediately carried over,
though an accurate calculation may require that the numerical 
techniques be more sophisticated, depending on the model chosen.
The other extension is to consider instead the Dirac equation,
in order to treat spin-1/2 particles instead of the spin-0 type
represented by the KG equation.  This would of course require
a new analysis in parallel with the analysis presented here,
except that the equations determining the metric from the
density profile would remain the same.

\appendix*   

\section{Reduction of the radial equation} \label{sec:Appendix}

In order to eliminate the first-order derivative from the radial
equation (\ref{eq:radialKG}), we write the radial wave function $R_l(r)=u_l(r)/h(r)$
in terms of a modified radial wave function $u_l$ and a function
$h(r)$ to be determined.  In the ordinary nonrelativistic Coulomb case
$h$ is known to be simply equal to $r$, but that would not be
sufficient here.  The term in (\ref{eq:radialKG}) that depends on
derivatives of $R_l$ is
\be
\frac{1}{r^2}\frac{d}{dr}\left(\frac{r^2}{g_{rr}}\frac{dR_l}{dr}\right)
= \frac{1}{g_{rr}}\frac{d^2R_l}{dr^2}+\left(\frac{2}{g_{rr}}-\frac{g'_{rr}}{g_{rr}^2}\right)\frac{dR_l}{dr}.
\ee
Substitution of $R_l(r)=u_l(r)/h(r)$ yields
\be
\frac{1}{r^2}\frac{d}{dr}\left(\frac{r^2}{g_{rr}}\frac{dR_l}{dr}\right)
=\frac{1}{hg_{rr}}\left\{\frac{d^2u_l}{dr^2}
  +\left[\frac{2}{r}-\frac{2h'}{h}-\frac{g'_{rr}}{g_{rr}}\right]\frac{du_l}{dr}
  -\left[\frac{h''}{h}-\frac{2(h')^2}{h^2}+\frac{2h'}{rh}
         -\frac{h'g'_{rr}}{hg_{rr}}\right]u_l\right\}.
\ee
The coefficient of $\frac{du_l}{dr}$ will be zero if $2hg_{rr}-2rh'g_{rr}-rhg'_{rr}=0$
or
\be
\frac{h'}{h}=\frac{1}{r}-\frac12\frac{d}{dr}\ln g_{rr}.
\ee
This can be directly integrated to obtain the logarithm of the solution, $h=r/\sqrt{g_{rr}}$.

\begin{acknowledgments}
R.D.L. gratefully acknowledges the support of a Research
Assistantship from the Department of Physics and Astronomy
at the University of Minnesota-Duluth.
\end{acknowledgments}

% If your manuscript is conditionally accepted, the editors will ask you to
% submit your editable LaTeX source file.  Before doing so, you should move
% all tables and figure captions to the end, as shown below.  Tables come 
% first, followed by figure captions (with figure inclusions commented-out).
% Figures should be submitted as separate files, collected with the
% LaTeX file into a single .zip archive.

%\newpage   % Start a new page for tables

%\newpage   % Start a new page for figure captions


\begin{thebibliography}{99}
\bibitem{units} We use units in which $\hbar=1$ and $c=1$, but not $G=1$.
\bibitem{positive} Only positive energy solutions are considered.  The negative
energies from the negative square root require field theory for their
interpretation in terms of antiparticles.
\end{thebibliography}
\end{document}